# Characterization of the large-size NDL EQR20 silicon photomultipliers

Yu. A. Melikyan [a,1], I.G. Bearden [b], V. Buchakchiev [c], S. Jia [b], V. Kozhuharov [c] and I. P. Møller [b]

[a] *Helsinki Institute of Physics, University of Helsinki, P.O. Box 64, FI-00014, Finland*
[b] *Niels Bohr Institute, University of Copenhagen, Jagtvej 155A, 2200 Copenhagen, Denmark*
[c] *Faculty of Physics, Sofia University "St.Kliment Ohridski", 5 J. Bourchier Blvd, 1164 Sofia, Bulgaria*

## Abstract

Unlike most commercially available silicon photomultipliers (SiPMs), EQR20 SiPMs produced by the Novel Device Laboratory (NDL) avoid using individual resistors to quench the avalanche multiplication of the microcells. Instead, bulk resistance of the epitaxial silicon layer is used, and the signal is directly collected at a common anode plane. This allows for the fabrication of SiPMs as large as 6.24 x 6.24 mm$^2$ while keeping the recovery time below $\tau$ = 25 ns. These devices can be composed of microcells with 20 μm pitch while reaching PDE above 50% and $10^6$ gain at 5 V overvoltage. On the other hand, a crosstalk level from 20% to 40% is observed for overvoltages from 3 V to 5 V. Moreover, significant pulse shape distortion is observed for pulses above ~100 pC, corresponding to microcell occupancy of a few percent. This work provides an independent determination of the performance parameters of the EQR20 11-6060D-S SiPMs. Influence of the pulse shape distortion is discussed from the perspective of the applicability of these devices to the scintillator-based forward hadron calorimeter (FoCal-H) of the ALICE apparatus at CERN.

## 1. Introduction

The Silicon Photomultiplier (SiPM), also known as Multi-Pixel Photon Counter (MPPC), is a modern photosensor for low-intensity light detection [1, 2]. Composed of thousands of individual micron-scale single-photon avalanche diodes (SPADs), which we refer to as microcells below, SiPMs produce an electronic signal proportional to the number of detected photons. The signal produced by one microcell is independent of the number of photons detected in it. Therefore, the proportionality between the number of detected photons and the signal charge deviates from linear once the Poisson probability of simultaneous detection of more than one photon in the same microcell becomes significant.

To extend the linear range of SiPM operation, a single SiPM must comprise a larger number of microcells. Reducing the microcell size has its limits, since it causes a reduction of the photon detection efficiency (PDE) and gain of the device [3, 4]. Substrates of larger area (e.g. 6x6 mm$^2$ or more) allow for SiPMs of larger number of microcells, thus increasing the useful dynamic range in terms of light intensity. Being directly proportional to the SiPM area, the capacitance grows significantly, extending the recovery time and the trailing edge of the output signal considerably [5], often exceeding $\tau$ = 100 ns.

A widespread SiPM production technology uses quenching resistors to connect each individual microcell with the power line [2]. Another technology relies on the bulk resistance of the epitaxial silicon layer to quench the avalanche multiplication process and recover the microcell [3, 5]. Below we refer to this technology as EQR (Epitaxial Quenching Resistance). Avalanche electrons are directly collected at a common anode on the device's surface. Minimizing the blind area at the SiPM input and reducing the terminal capacitance, this technology allows the production of large-size SiPMs featuring much higher PDE and much faster recovery than the current widespread production technology [6].

[1] Corresponding author. E-mail: yumelikyan@gmail.com

Figure 1 clearly illustrates the difference in response of a pair of otherwise similar SiPMs produced according to the two different technologies.

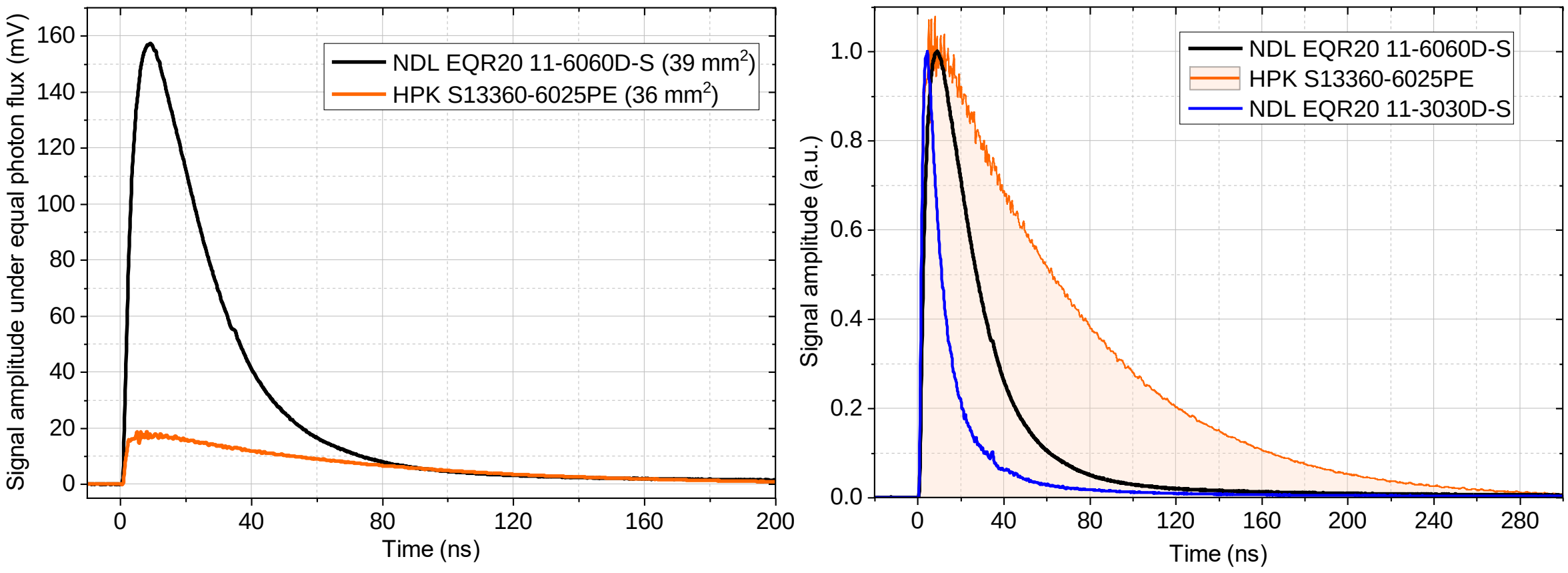

Figure 1. Left - response of an NDL EQR20 11-6060D-S SiPM (39 mm$^2$) and a Hamamatsu S13360-6025PE SiPM with individual quenching resistors (36 mm$^2$) to a sub-ns flash of 440 nm light. The pulse shapes were averaged over 2000 waveforms measured at 5 V overvoltage under an equal photon flux of 25 photons / mm$^2$ / pulse. The mismatch in the pulses' height and width are due to the different PDE, gain, crosstalk, sensitive area and RC of the devices. Right – same data reproduced with each pulse normalized to its peak amplitude. The blue signal is from a smaller NDL EQR20 11-6060D-S SiPM, sized 3x3 mm$^2$, measured under the same photon flux at 4.7 V overvoltage.

The advantages of the EQR technology for the photon readout in a "spaghetti-type" forward hadron calorimeter (FoCal-H [7]) of the ALICE experiment at HL-LHC are:

- A combination of the high PDE and high number of microcells (and thus the large dynamic range) can provide freedom to choose the number of scintillation fibers read out by a single SiPM, thus optimizing the detector granularity.
- A shorter pulse width and faster recovery can extend the operability of a SiPM when its dark count rate is increased as a result of radiation damage.
- Faster recovery also extends the rate capability of the device, particularly relevant given the 25 ns spacing between the two adjacent bunches of HL-LHC [8].

EQR SiPMs with 20 μm microcell size and sensitive area of up to 39 mm$^2$ are commercially available from Novel Device Laboratory (NDL EQR20 11-6060D-S) [6]. This combination of the sensitive area and the microcell size is optimal for their potential use in FoCal-H [7]. The scarce data available in literature on the performance of this SiPM type did not allow us to determine if they indeed outperform the SiPMs produced according to the widespread technology in our application. Therefore, we have conducted a dedicated characterization of the NDL EQR20 11-6060D-S SiPMs, and the results are provided here to facilitate the potential use of the NDL EQR SiPMs in other applications.

## 2. The experimental set up and measurement program

The schematic of the experimental setup is provided in Fig. 2. Two or more SiPMs were tested simultaneously in a light-tight box. For the dark noise measurements, an external light absorber was added around the SiPM input window, thus ensuring no photons scattered from inside one photosensor under study get detected in the others. Those measurements related to external photon detection were done with pulsed light produced by a picosecond laser monitored by a reference photosensor (Planacon XP85002/FIT-Q PMT [9]). We used a Picoquant PDL800-B laser with an LDH-P-C-440M laser head. It features 440 nm ± 10 nm wavelength, ≤20 ps jitter and ≤55 ps pulse width [10].

The laser beam was defocused by an optical fiber coupler and passed through an optical attenuator. With ±13° aperture of the multimode fiber used, output light was projected to a diffuse white reflector, ensuring a spot large enough to provide a homogeneous photon flux distribution across the sensitive area of all photosensors used in the set up. The spread in the photon flux across all photosensors was <1% - see [9] for the detailed description and results of the homogeneity measurement. This allowed us not only to test different SiPMs under the same photon flux but also to measure its absolute value thanks to the known sensitive area, quantum efficiency, collection efficiency and gain of the reference PMT [9].

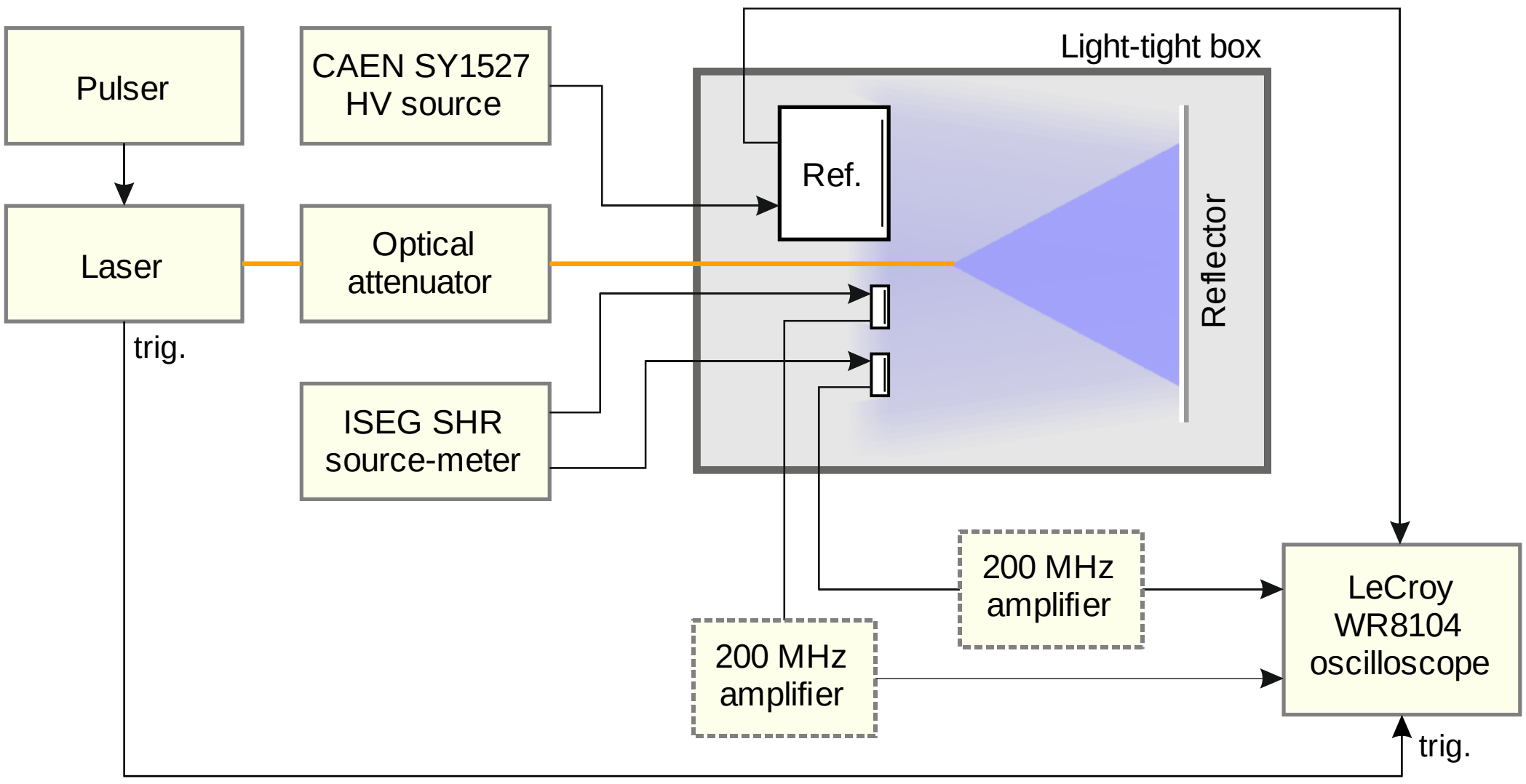

Figure 2. Schematic of the experimental set up. The 200 MHz amplifiers were used in a subset of measurements.

Positive bias voltage was supplied to SiPM cathode using the basic connection circuit recommended by the manufacturer – see Fig. 3.

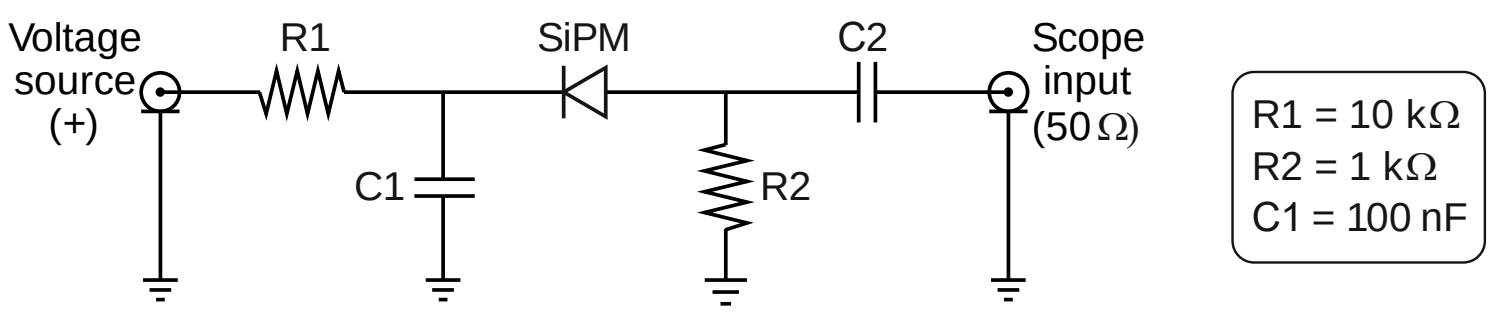

Figure 3. The simplest SiPM connection diagram recommended by the manufacturer was used [6]. R1, R2, and C1 values were kept constant in all cases, while the coupling capacitance C2 was shorted or altered in different measurements, as reflected in the text.

The measurement program was intended to verify and complement those NDL specification parameters relevant to the possible SiPM application for scintillation photon readout of the future ALICE FoCal hadron calorimeter [7]:

- breakdown voltage defined from bias I-V curves and from relative gain measurement;
- pulse shape stability at various overvoltages;
- effect of the optical crosstalk between microcells derived from the dark noise spectra;
- visible value of the dark count rate (DCR) and the true DCR value derived based on the actual pulse shape;
- absolute PDE corrected for the crosstalk value;
- response linearity and pulse shape stability versus absolute flux of the incoming photons;
- response as a function of the time delay between two consecutive pulses and their intensity.

## 3. Results

### *3.1. Breakdown voltage*

SiPM breakdown voltage at 26 ± 1 °C was determined using two independent experimental techniques. First, the bias voltage of tested SiPMs was increased at a 0.05 V/s rate, which is slow enough to spot the rise in bias current characteristic for the region of avalanche multiplication with $V_{bias} > V_{br}$. The I-V curves measured for two tested SiPMs of the same type are shown in Fig. 4.

The standard deviation of the leakage current is $\sigma(I_{leakage}) \approx 3$ nA. We determined $V_{br}$ as the visible starting point of the rise of dark current once $I_{dark}$ exceeds $2 * \sigma(I_{leakage})$. The result is $V_{br} = 27.4$ V for both tested SiPMs, in agreement with the typical $V_{br} = 27.5$ V outlined in the datasheet [6].

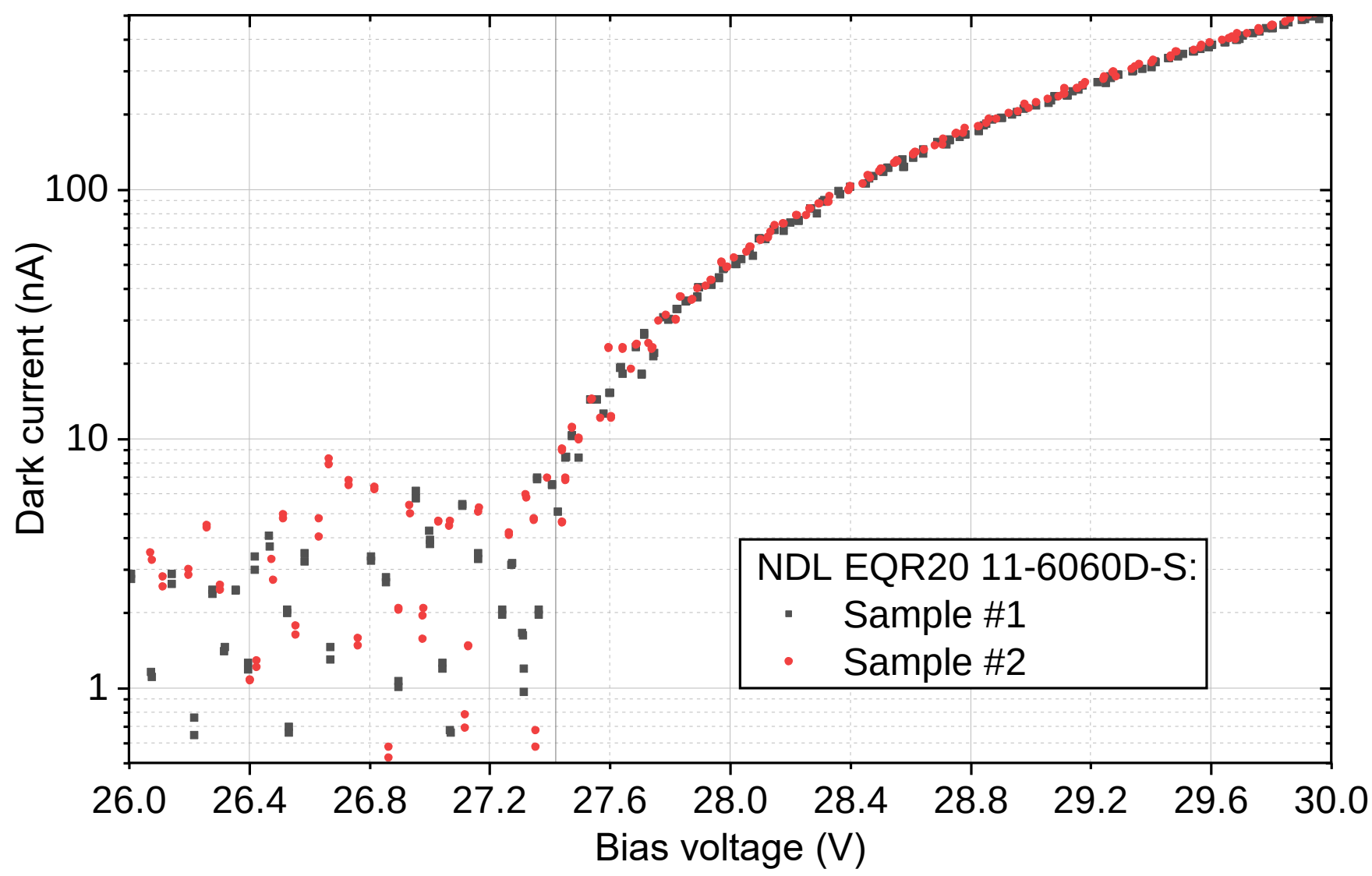

Figure 4. Dark current dependence on the bias voltage for two NDL EQR20 11-6060D-S SiPMs as measured by the ISEG SHR 4260 source-meter at 0.05 V/s ramp speed.

The $V_{br}$ determination result was cross-checked independently for NDL SiPM #1 by measuring the charge spectra of the dark signals passed via an external 200 MHz LeCroy 612M amplifier. Typical spectra measured above the detection threshold of 0.5 photoelectrons (p.e.) at various overvoltages are shown in Fig. 5.

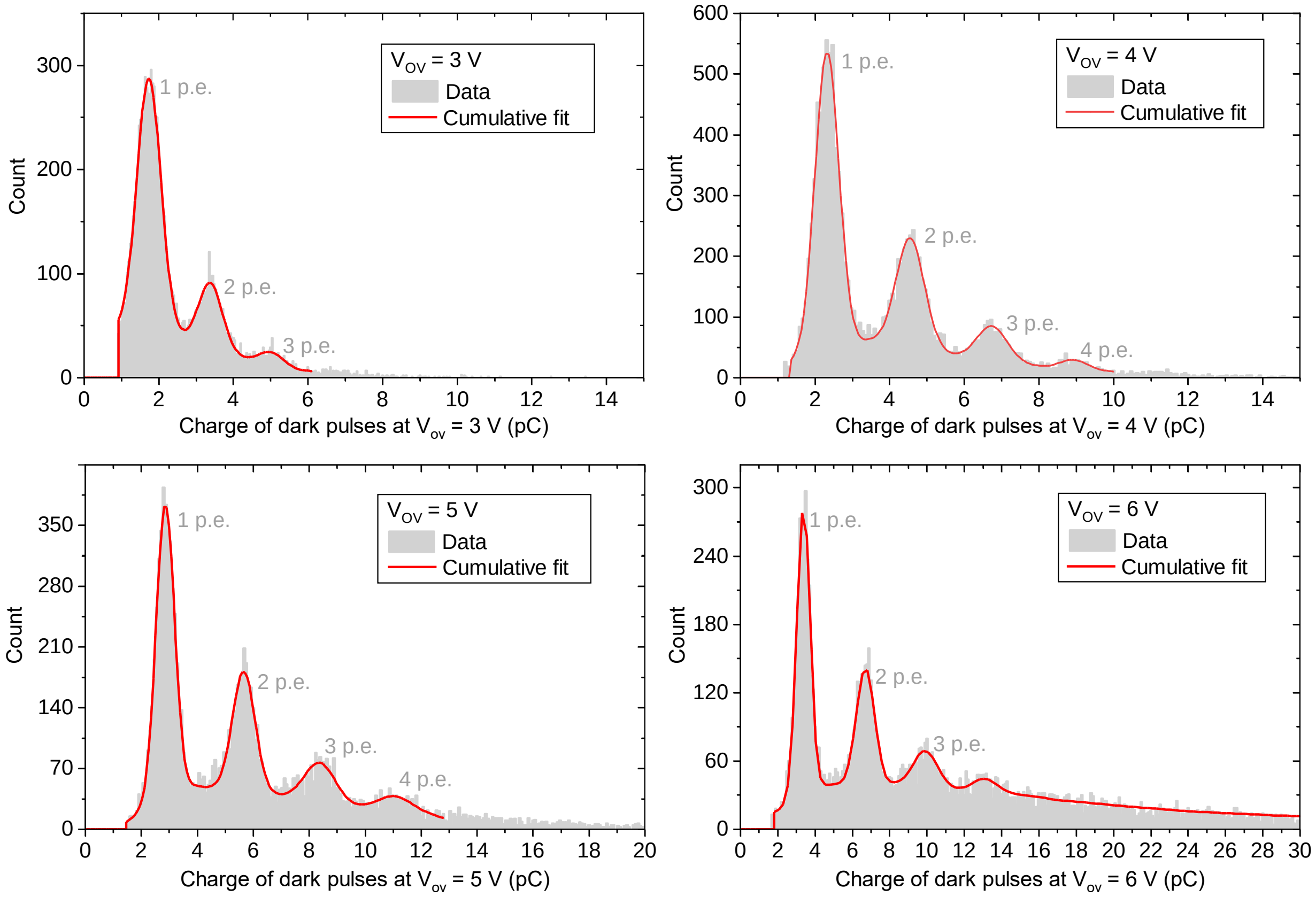

Figure 5. Charge spectra of dark pulses of an NDL SiPM sized 6.24 x 6.24 mm$^2$ measured at overvoltages from 3 V to 6 V. An external amplifier was used to better distinguish the discrete peaks at low overvoltage.

A combined fit of multi-gaussian peaks with a common baseline function was used to define the positions of individual peaks [11]. The single photoelectron (SPE) charge was defined as the difference in the fitted means of 2 p.e. and 1 p.e. peaks. The linear nature of the SPE charge dependence on the overvoltage was used to define the breakdown voltage, as shown in Figure 6. The result, Vbr = 27.42 V ± 0.04 V, agrees perfectly with the result of the $V_{br}$ determination from the IV curve described above.

### *3.2. Absolute gain*

The dependence of the absolute gain on the SiPM overvoltage can be derived from the data shown in Fig. 6 by correcting the vertical scale for the gain of the external amplifier. However, to avoid the possible influence of the amplifier's integral and differential nonlinearities on the calculation result, we measured the SPE charge spectra of low-intensity laser pulses at $V_{ov} \approx 5$ V directly, bypassing the amplifier. With pulses as narrow as 1.5 mV/pC (see Fig. 1), the large-size NDL SiPM feature sufficient SPE resolution to distinguish the discrete structure of the pedestal, 1 p.e. and 2 p.e. peaks.

The absolute SPE charge measured at three different bias voltages is shown in Fig. 7, along with the $V_{br}$ value defined in Section 3.1. The linear fit of the four available points shows the dependence of the absolute gain (G) on the overvoltage. As can be seen from the fit, the NDL EQR20 11-6060D-S SiPM features absolute gain as high as 5.7*10$^5$ at $V_{ov}$ = 3 V, almost reaching the gain of 10$^6$ at $V_{ov}$ = 5 V. Given the relatively small microcell pitch of 20 μm, this value is competitive with SiPMs having individual quenching resistors [4, 12].

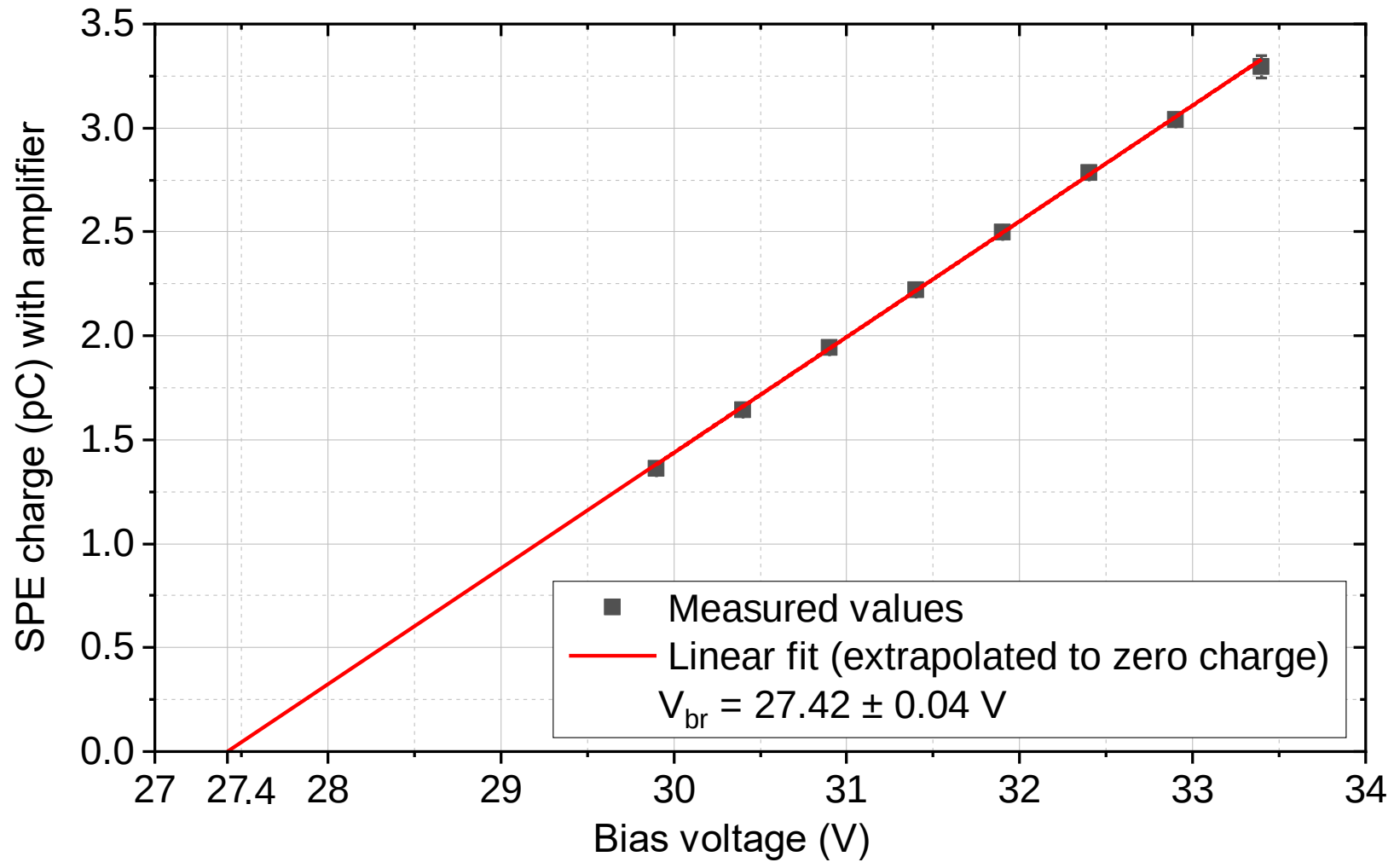

Figure 6. Determination of the breakdown voltage of the NDL SiPM sized 6.24 x 6.24 $mm^2$ from the SPE charge dependence on the bias voltage.

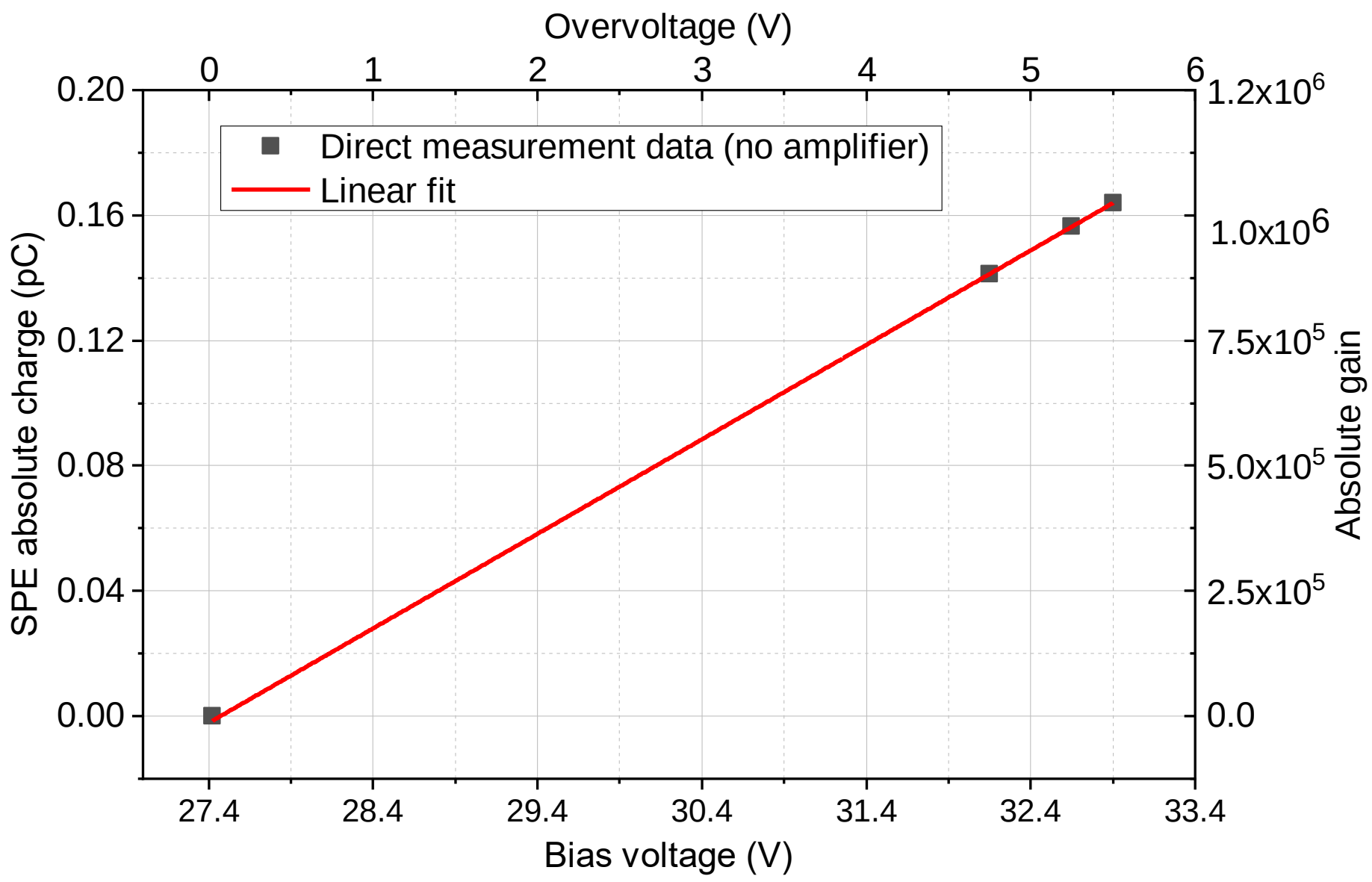

Figure 7. The mean value of the SPE charge measured without the external amplifier. Error bars are smaller than the data point symbols. Having the $V_{br}$ value known, the linear fit of the combined data represents the dependence of absolute gain (indicated on the right vertical axis) of NDL EQR20 11-6060D-S SiPM versus overvoltage.

### *3.3. Pulse shape stability versus overvoltage*

NDL EQR20 11-6060D-S SiPMs show some variation of the pulse shape at different overvoltages measured with the same incident light intensity. Pulses from low-intensity laser illumination (~0.2% microcells occupancy at $V_{ov}$ = 3 V) become shorter while increasing $V_{ov}$ from 2.5 V up to 5 V. At even larger overvoltage, the pulses become longer – see Fig. 8. Detailed quantitative parameters of the rise time ($t_{rise}$, the time difference between signal crossings at relative amplitude thresholds) and the fall time ($\tau_{fall}$, the characteristic time of the trailing edge fitted with an exponential function) are provided in Fig. 9 as a function of the overvoltage.

When using the DC-coupled SiPM circuit (C2 shorted, as in Fig.9 top-left and bottom), the $t_{rise}$ = 5.3 ns at $V_{ov}$ = 2.5 V, dropping to $t_{rise}$ = 4.4 ns just below $V_{ov}$ = 5 V; above that, the rise time grows again, reaching $t_{rise}$ = 5.6 ns at $V_{ov}$ = 6 V. Fall time changes from $\tau_{fall}$ = 28 ns at 2.5 V overvoltage, dropping to 22 ns before increasing to 34 ns at $V_{ov}$ = 6 V. This effect could be mitigated by using a coupling capacitor as a high-pass filter. The dependence shown in Fig. 9 (top-right) is almost flat thanks to the use of C2 = 62 pF as the coupling capacitor.

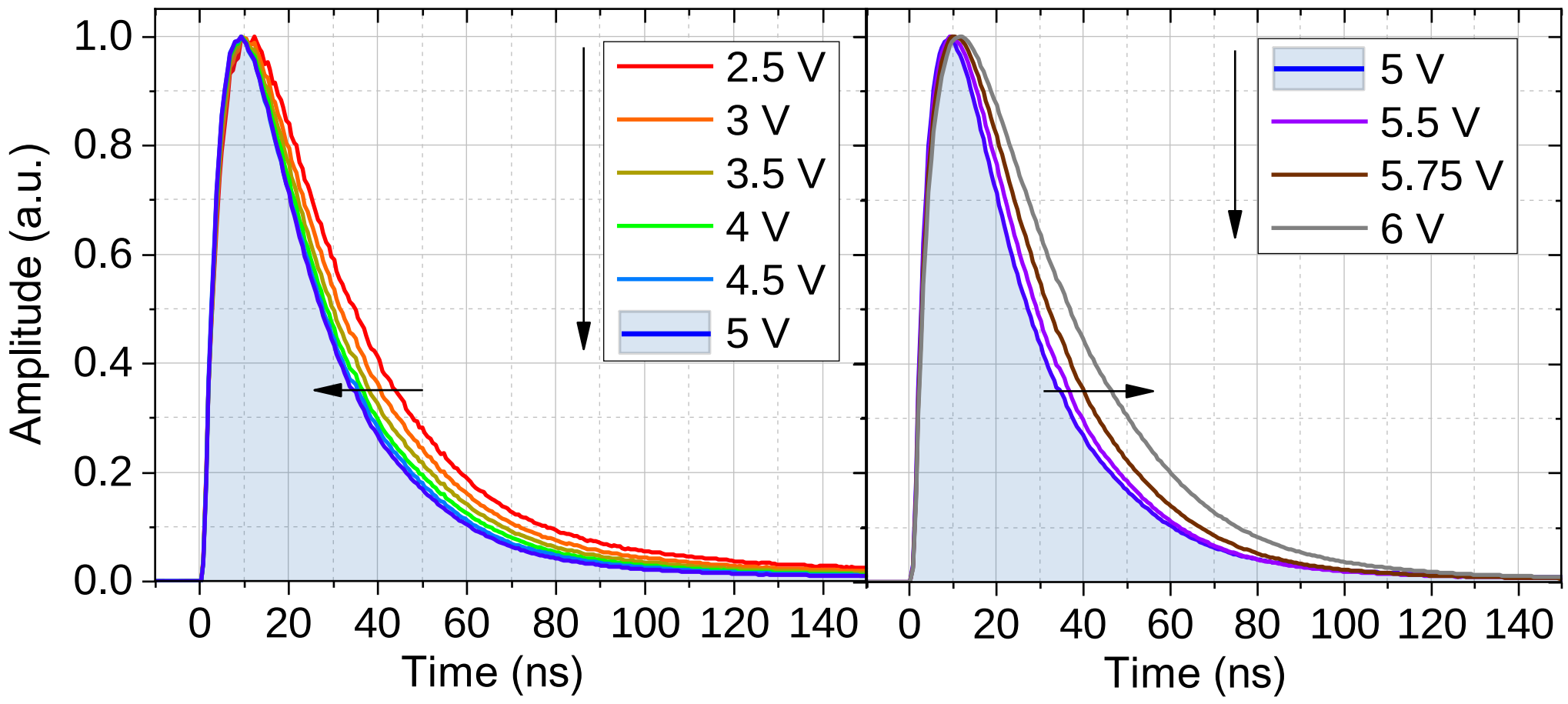

Figure 8. Pulse shapes of the same NDL EQR20 11-6060D-S SiPM under picosecond laser illumination at different overvoltages. For overvoltages from 2.5 to 5 V, the pulse narrows and then broadens for overvoltages above 5 V. Horizontal arrows highlight the different trend of the change of the pulse width as overvoltage grows.

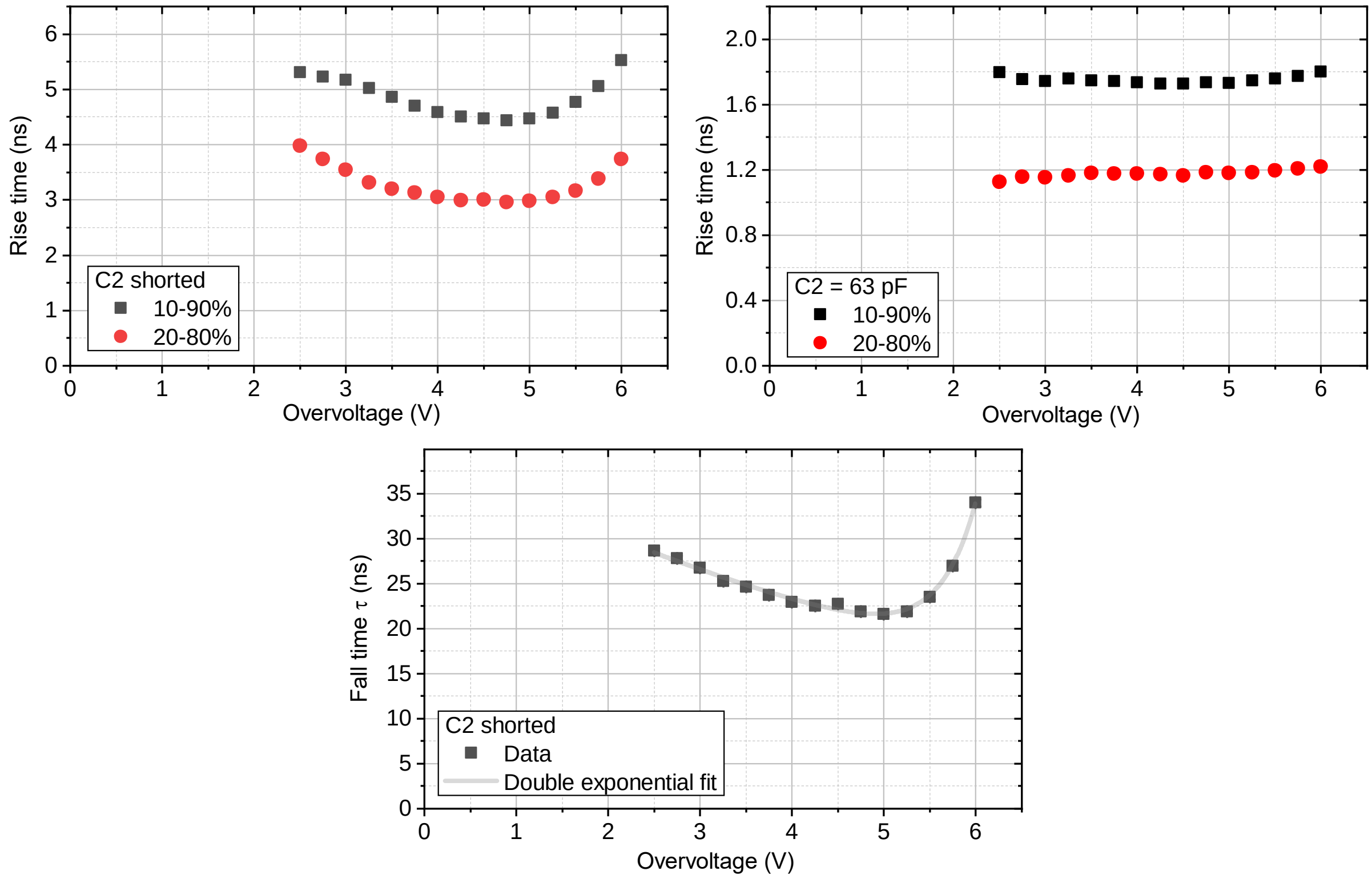

Figure 9. Top: interpolated rise time of an NDL EQR20 11-6060D-S SiPM sized 6.24 x 6.24 $mm^2$ under low-intensity picosecond laser illumination. The left graph is for the case when no coupling capacitance is used (C2, see Fig. 3). The right graph was measured with C2 = 63 pF. Bottom: characteristic time of the trailing edge fitted with an exponent function ($\tau_{fall}$). The resulting dependence of $\tau_{fall}$ versus $V_{ov}$ can be fit with a double exponential (adj. R-square = 0.982).

*3.4. Dark count rate and optical crosstalk*

Operating at room temperature, large-area SiPMs suffer from high dark count rates (DCR) on the order of MHz. Given also the significant pulse width (FWHM > 30 ns), the probability of two uncorrelated dark pulses randomly overlapping in time becomes significant. This affects the visible value of DCR determined using a simple threshold (e.g., 0.5 p.e.). It also affects the visible value of prompt crosstalk (CT) level defined according to the standard technique (number of dark pulses with amplitude above 1.5 p.e. divided by the number of dark pulses with amplitude above 0.5 p.e. [13]). To obtain the true value of DCR and the true value of CT, we measure their "visible" values and use a simplified Monte-Carlo simulation to extract the true values from the visible ones.

In our technique of measuring the visible DCR, the first dark pulse is detected once its amplitude exceeds 0.5 p.e. in a single rise. The counter is then deactivated until the waveform crosses the 0.5 p.e. threshold again in the opposite direction. Regardless of the total pulse amplitude (i.e. even in case of a statistical pile-up), only one dark pulse is counted until the waveform falls below 0.5 p.e. Afterwards, another dark pulse is counted only if the waveform both exceeds the 0.5 p.e. threshold again and has a rise of at least 0.5 p.e. in a single event.

The simulated curve used to reconstruct the true DCR value from the measured ("visible") one is shown in blue in Fig. 10. The dashed red line shows another derivative of the same simulation – the ratio of the purely statistical overlap of two or more pulses (resulting in signal amplitudes above 1.5 p.e.) to the visible DCR. By definition, the latter parameter does not include any correlated noise, but only a random overlap in time of multiple dark pulses. Therefore, subtracting the simulated statistical contribution from the visible CT value results in the true CT.

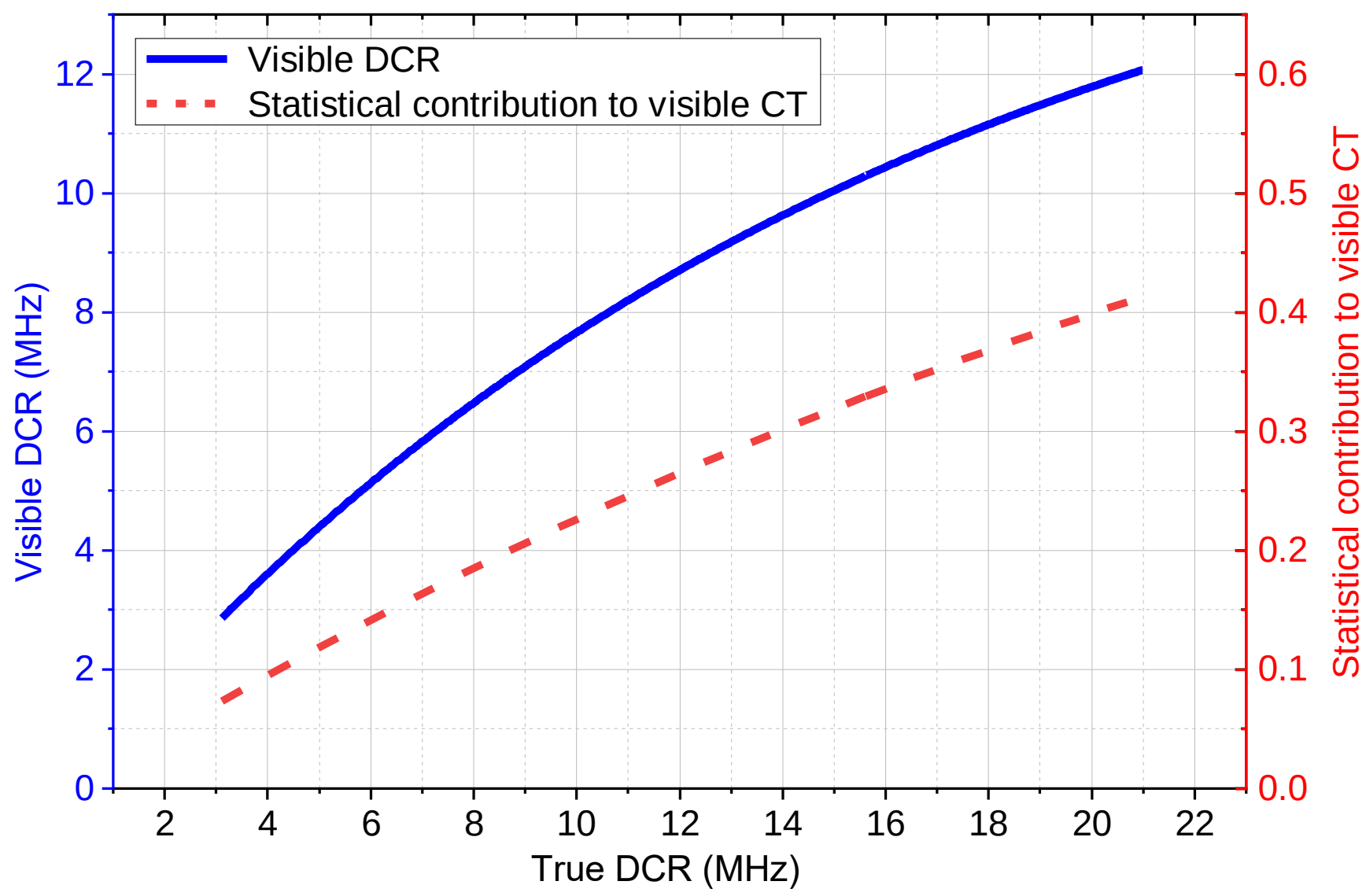

Figure 10. Statistical corrections to the visible DCR and visible CT values simulated for the pulse shapes of the NDL EQR20 11-6060D-S SiPM sized 6.24 x 6.24 $mm^2$.

Figure 11 represents in black the "raw" results on the DCR and CT as measured according to the simplistic techniques: fixed 0.5 p.e. threshold for the DCR and a "blind" ratio of the number of dark counts with 1.5+ p.e. and 0.5+ p.e. amplitude. The points of Fig. 11 marked with red circles are based on the same experimental data but corrected for the simulated dependencies from Fig. 10, thus representing the true DCR and prompt crosstalk.

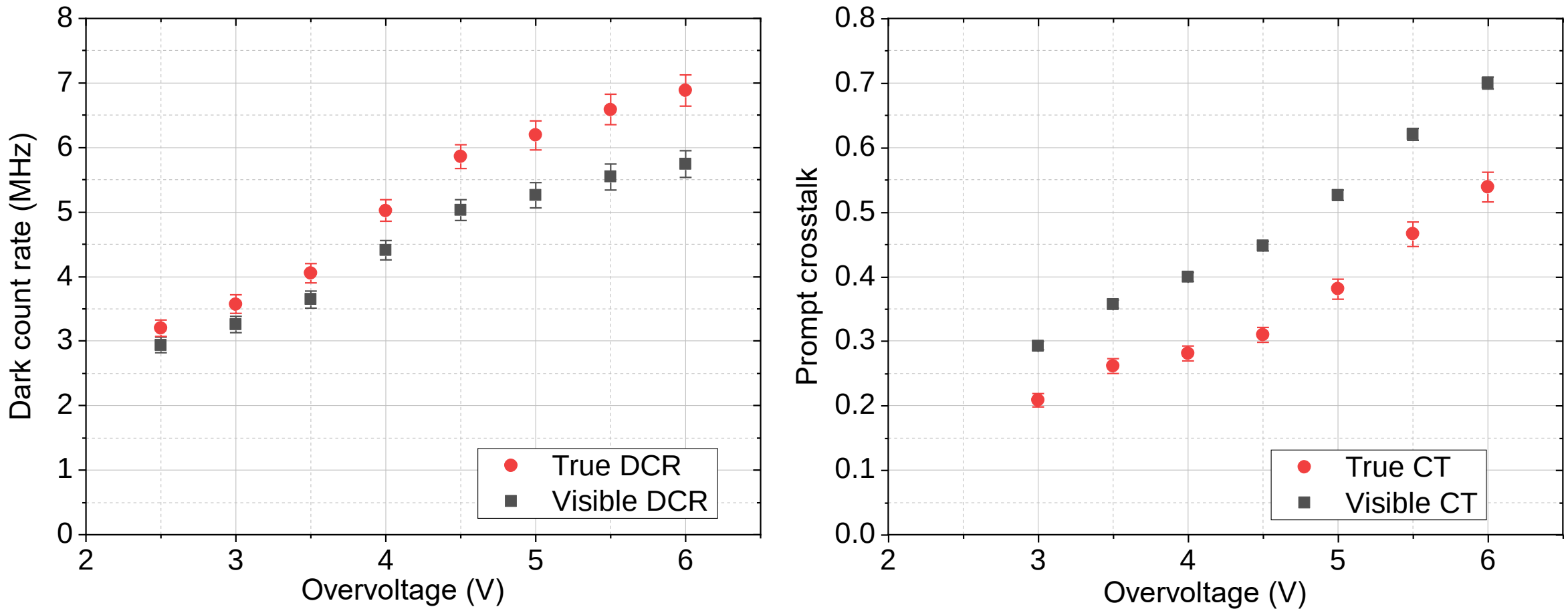

Figure 11. Dark count rate and prompt crosstalk measured at 25°C as a function of the overvoltage of the NDL EQR20 11-6060D-S SiPM sized 6.24x6.24 $mm^2$.

### *3.5. Photon detection efficiency*

Absolute photon detection efficiency (PDE) was measured under pulsed light ($\lambda$ = 440 nm) as a ratio between the signal charge detected by the NDL SiPM and the photon flux detected by the reference PMT. The charge of the NDL signal was counted in p.e. using the absolute gain from Fig.7 and corrected for the crosstalk shown in Fig.11. Photon flux was uniformly distributed across the area of the reference PMT and tested SiPM (variation of the light distribution <1%, see Section 2 in [9]). It allowed to derive the number of photons per pulse per area hitting the SiPM thanks to the precisely known parameters of the Planacon PMT. The result is shown in Figure 12 – NDL EQR20 11-6060D-S SiPM features remarkably high PDE for 440 nm light of 47 ± 3 (%) at $V_{ov}$ = 3 V and 55 ± 4 (%) at $V_{ov}$ = 5 V.

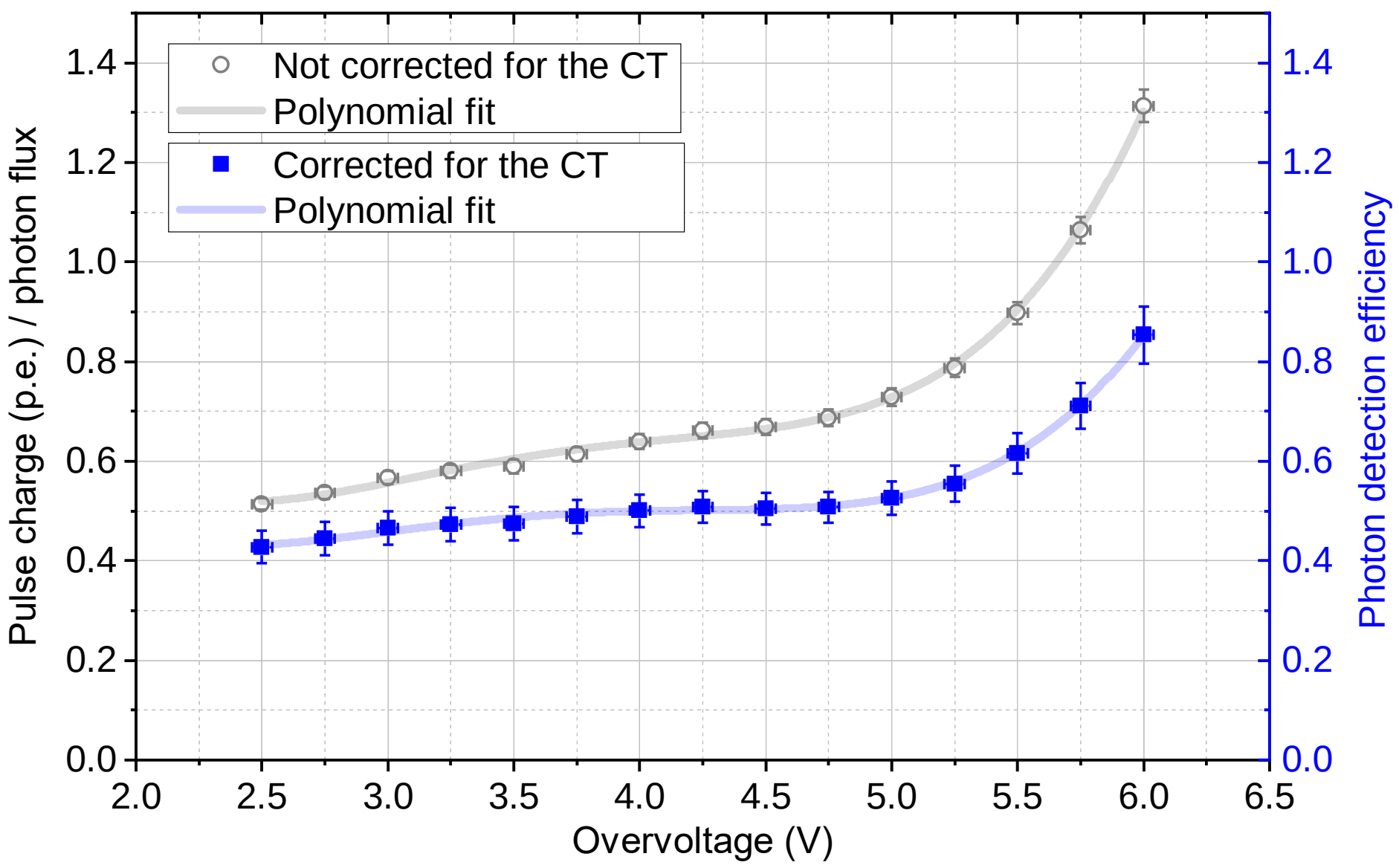

Figure 12. Photon detection efficiency of the NDL EQR20 11-6060D-S SiPM sized 6.24 x 6.24 $mm^2$. The gray circles show the uncorrected data, while the blue squares show the data corrected for cross talk as explained in the text. $4^{th}$ order polynomial fits are used to guide the eye.

*3.6. Dynamic range and linearity*

The saturation curve was measured for the NDL EQR20 11-6060D-S SiPM, and it was compared to the one for the Hamamatsu S13360-6025PE SiPM placed in the same laser light spot. Both SiPMs had similar sensitive area, but different pitch and number of microcells (NDL: 20 µm pitch, 39 mm$^2$ area, 97344 microcells; Hamamatsu: 25 µm pitch, 36 mm$^2$ area, 57600 microcells). The results are shown in Fig.13.

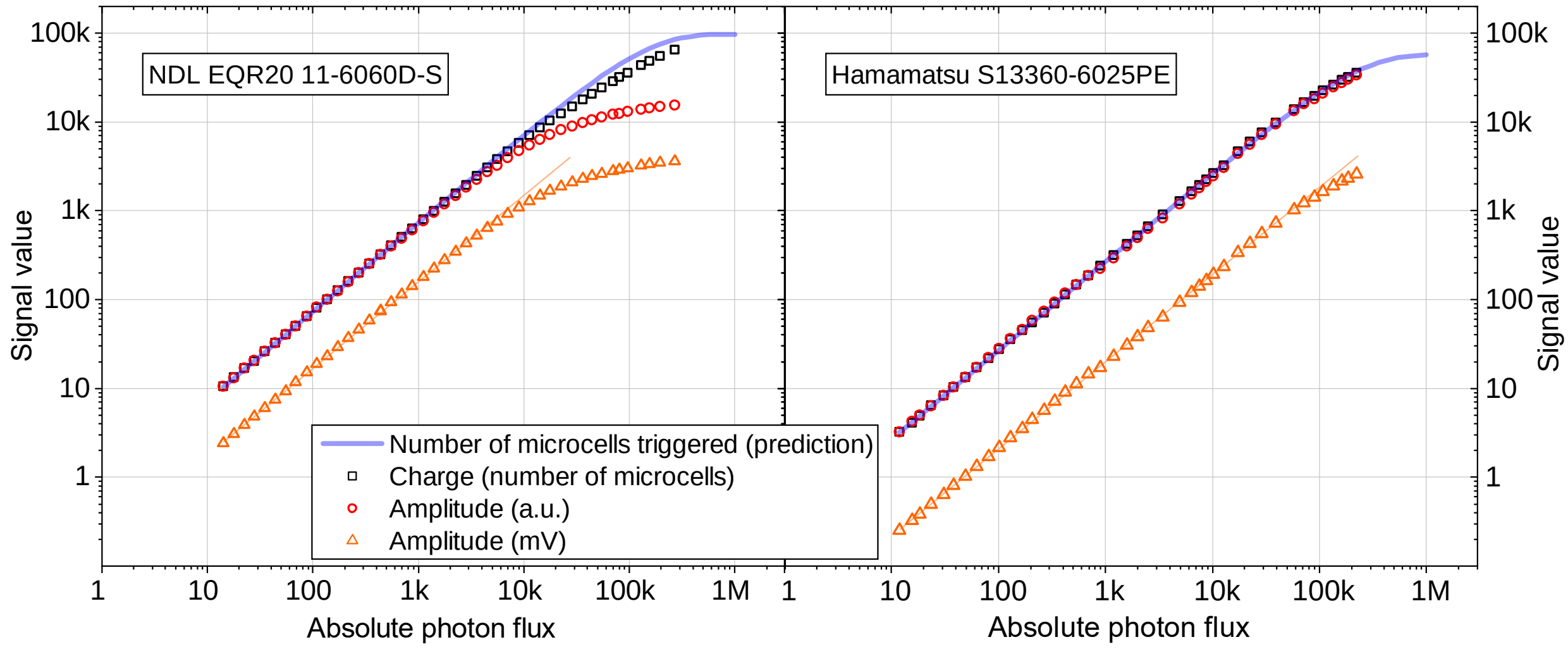

Figure 13. Output signal intensity integrated over a 600 ns-wide time window versus input light intensity for the NDL EQR20 11-6060D-S SiPM with 97k microcells and Hamamatsu S13360-6025PE SiPM with 57k microcells. For both devices, $V_{ov}$ = 5.0 V was used and the coupling capacitance C2 was shorted.

Note the predictive curve is calculated independently of the output pulse measurement. It is a pure Poisson expectation from the photon flux measured in absolute units corrected for the PDE and CT values from the independent measurements reported above for the NDL EQR20 11-6060D-S SiPM. Therefore, Fig.13 provides also a cross-check of those values. The same curve for the Hamamatsu S13360-6025PE SiPM is calculated from the specification data available from the manufacturer [4].

The Hamamatsu SiPM saturation curves both in terms of signal amplitude and charge behave in line with the Poisson prediction – experimental data deviates by less than 5% throughout the entire intensity range (up to 60% microcell occupancy reached in this measurement). The NDL curves deviate by more than 5% at microcell occupancies above 1% and 3% for amplitude and charge dependencies respectively. At 60% microcell occupancy, the amplitude deviates down to 23% of the predicted value, charge – down to 72%.

The significant deviations of the NDL EQR20 11-6060D-S saturation curves above 1% microcell occupancy at $V_{ov}$ = 5 V are a sign of their pulse shape distortion. Fig. 14 provides the actual waveforms captured at various photon fluxes. The quantitative details are summarized in Fig. 15. As can be seen from Fig.14 and 15, larger signals feature shorter rise time and extended fall time in the NDL EQR20 11-6060D-S SiPMs, while the signals of Hamamatsu S13360-6025PE remain stable. Fall time $\tau_{fall}$ of the NDL pulse exceeds 100 ns at larger occupancies, thus reducing the signal charge from the expected value as long as it is integrated over a limited time window (600 ns in this case). A similar effect characteristic for the NDL EQR SiPMs of other types was reported [5]. The mismatch between the experimental and simulated dynamic range of the 6 µm-pitch NDL SiPM reported in [14] is likely a consequence of the same effect.

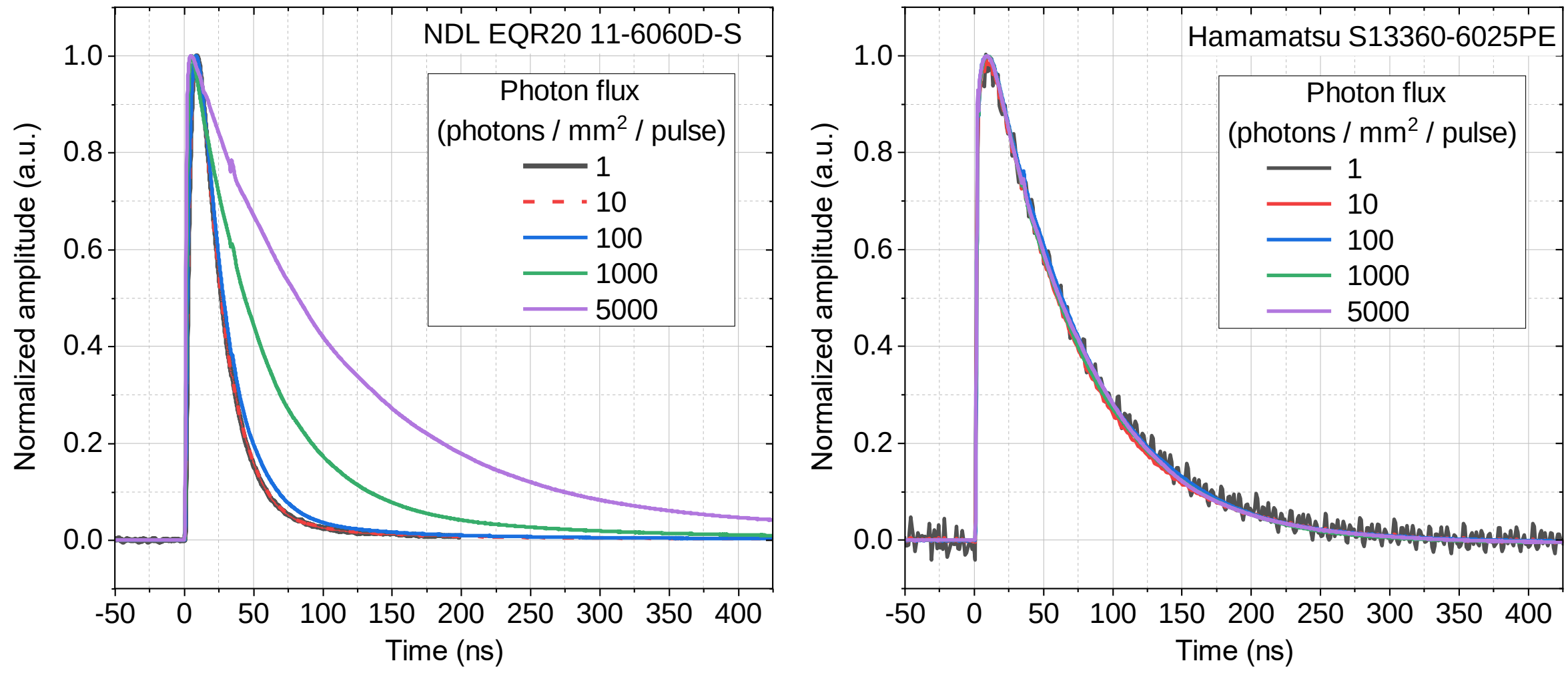

Figure 14. Waveforms of SiPM signals under picosecond laser illumination of various intensity.

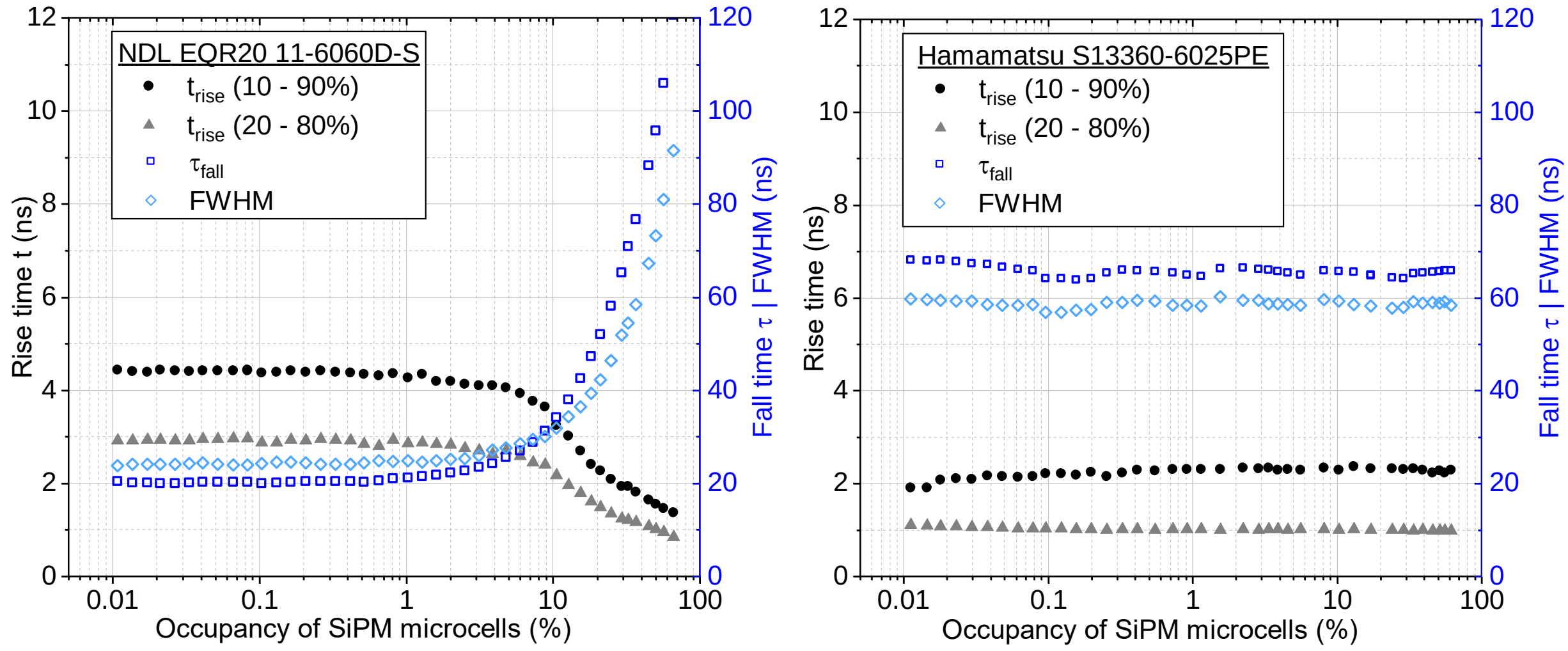

Figure 15. Rise time ($t_{rise}$), fall time ($\tau_{fall}$) and interpolated FWHM of signals of the NDL and Hamamatsu SiPMs as a function of the microcell occupancy.

While operated with the same microcell occupancy but under different overvoltage, output charge of a SiPM is different. Therefore, same measurements conducted at different overvoltages can reveal if the pulse shape distortion depends on the output charge, or on the microcell occupancy. Fig.16 provides the dependencies of the key pulse shape parameters both on the microcell occupancy and on signal charge, measured at different overvoltage. The two parameters – rise time $t_{rise}$ (10 - 90%) and fall time $\tau_{fall}$ – change with overvoltage (see Fig.9). Thus, for a proper comparison, timing parameters shown in Fig.16 are presented in arbitrary units, being normalized to yet undistorted values at low pulse charge and occupancy.

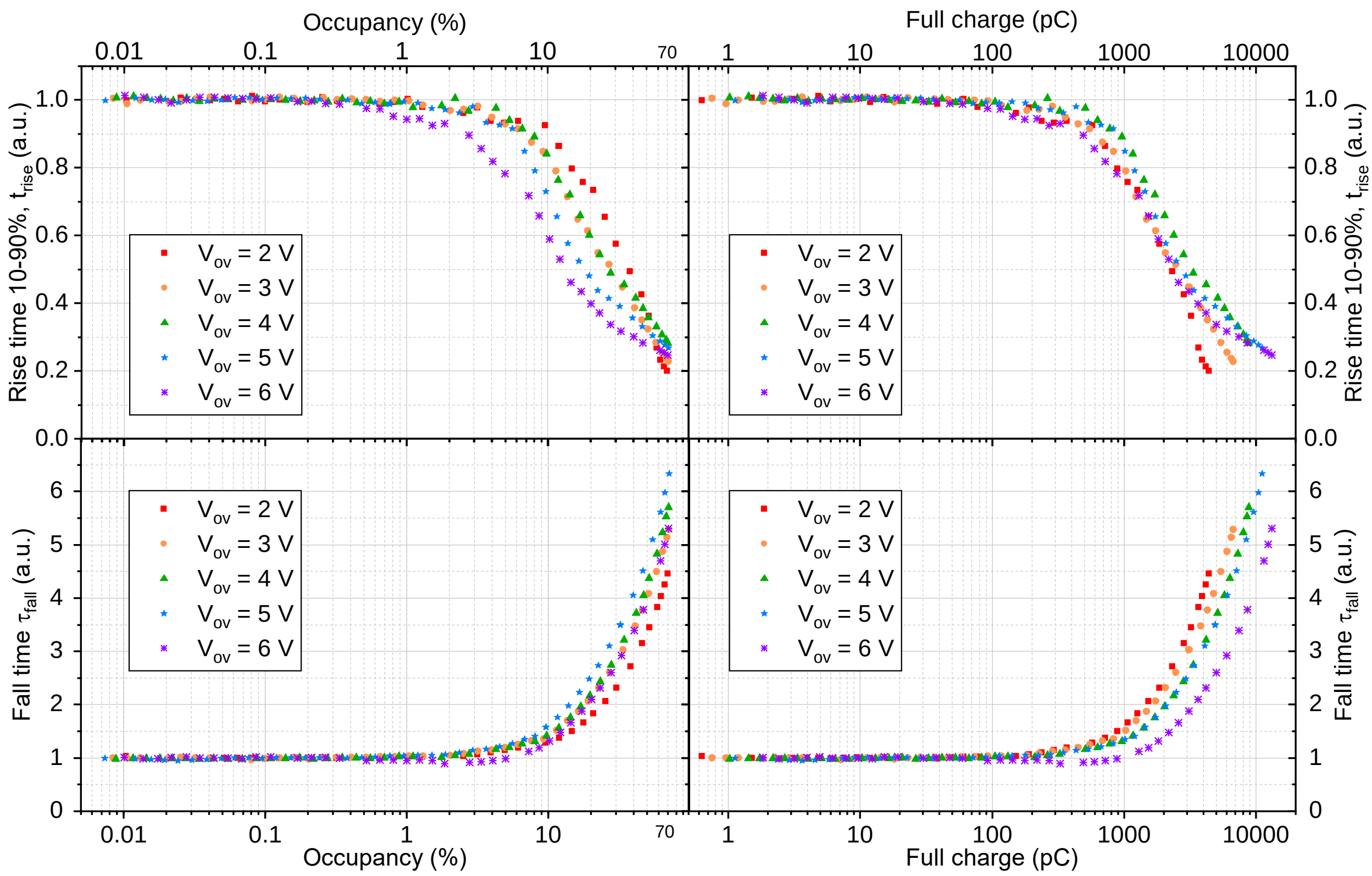

Figure 16. The interpolated 10 – 90% rise time $t_{rise}$ (top) and the exponent-fitted fall time $\tau_{fall}$ (bottom) of signals of the NDL EQR20 11-6060D-S SiPMs at various overvoltage. Same data is reproduced as a function of the microcell occupancy (left) and as a function of the signal charge (right) integrated over a 1.6 µs-wide time window.

### *3.7. Rate capability*

The rate capability of the NDL EQR20 11-6060D-S SiPM was studied from the perspective of their possible application to the future hadron calorimeter for the ALICE upgrade at HL-LHC. We estimated variation in the SiPM response under bursts of two pulses coming at a low rate (100 Hz) but with very short spacing (delay down to 25 ns). This technique can help to estimate the limits of scintillation fiber grouping to the same SiPM and also make a comparative study between two different SiPMs. We, therefore, simultaneously measured the performance of the NDL EQR20 11-6060D-S and Hamamatsu S13360-6025PE SiPMs. Variation of the SiPM response for the second pulse in a burst dependent on delay between the two pulses in a burst was measured at four different light intensities (from 10 photons / $mm^2$ / pulse to 3800 photons / $mm^2$ / pulse). The results are shown in Fig.17 as four plots, each with a pair of curves measured at the same pulse intensity.

A fair comparison of the performance of the NDL and Hamamatsu SiPMs based on Fig.17 is complicated due to the very different PDE and CT values of the two SiPM types. The experimental data was therefore rearranged in Fig.18 to reflect the difference in the number of microcells fired under the same pulse intensity. The shaded areas for the two SiPM types overlap, highlighting their similar performance: the NDL advantage of a larger number of microcells is counterbalanced by the drastic increase of $\tau_{fall}$ (and thus the recovery time) at larger pixel occupancies.

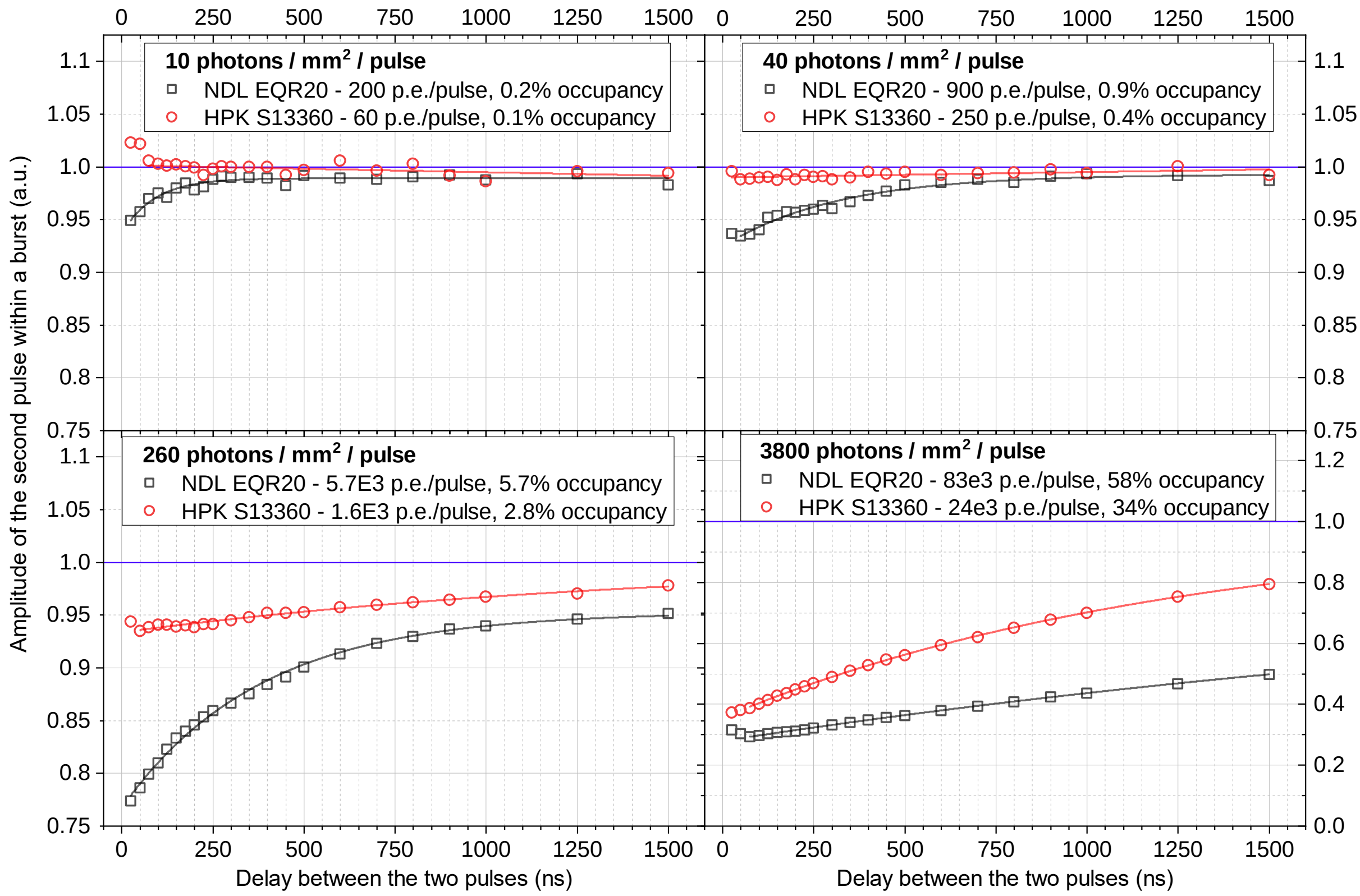

Figure 17. SiPM response as a function of the time delay between two consecutive pulses and their light intensity.

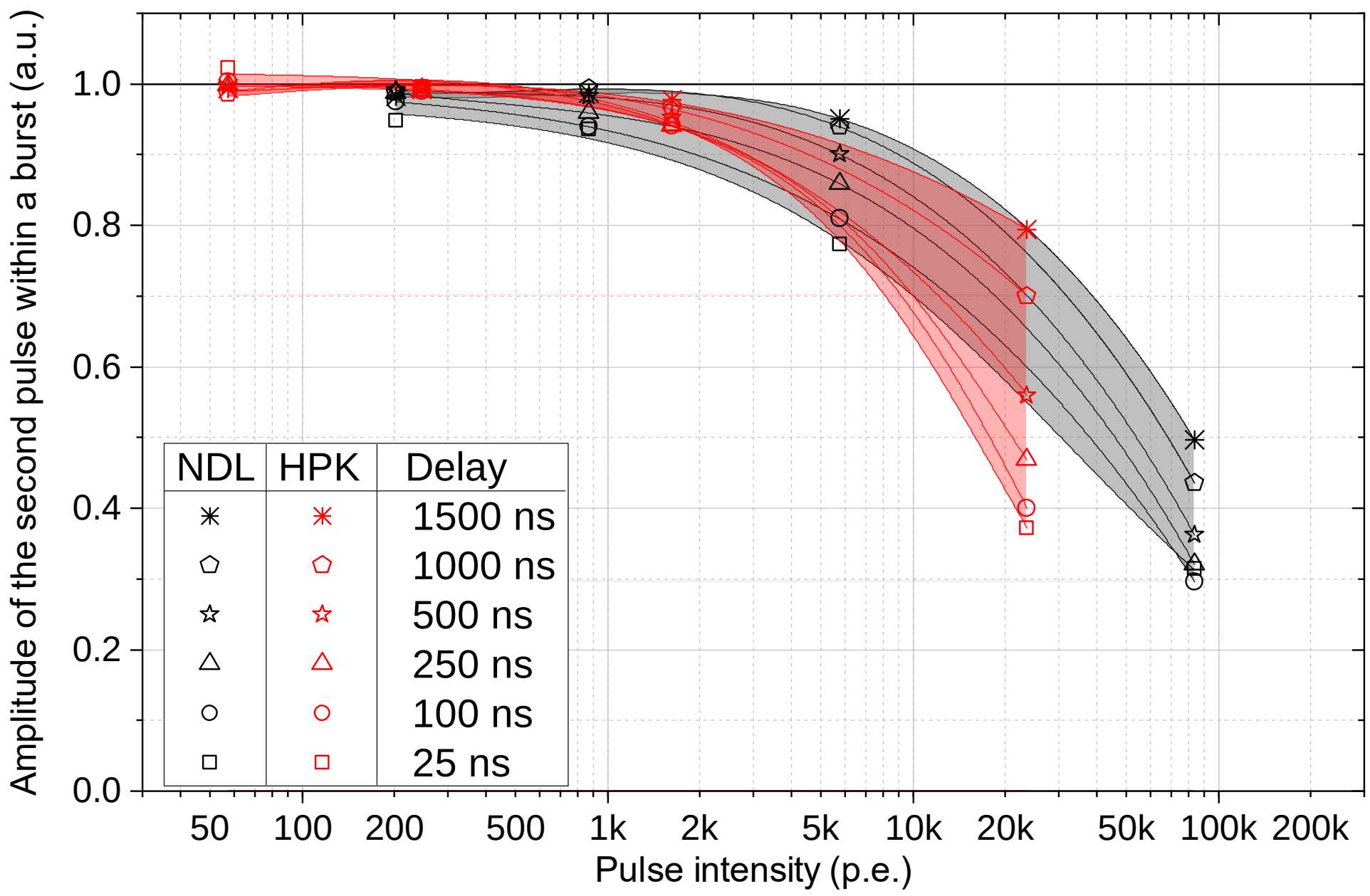

Figure 18. SiPM response as a function of the time delay between two consecutive pulses and their intensity in terms of the pulse intensity, i.e. the number of microcells fired.

## 4. Discussion

The distortion effect reported for the NDL EQR20 11-6060D-S SiPMs shall be distinguished between the deviations of the rise time and the fall time. The nature of the latter is discussed in [5] from the perspective of the recovery time variation dependence on the number of pixels fired. According to [5], it is caused by a combination of capacitive effects and signal propagation delays across the device. When more microcells are fired, the simultaneous recovery and recharging of multiple microcells cause pulse broadening. This broadening results from a transmission line effect, where signals from different microcells have varying delays before reaching the electrode. However, the trends of the timing parameters shown in Fig. 15 are flat up to ~1% occupancy. Also, we did not see any significant difference in $\tau_{fall}$ measured using picosecond laser illumination (~55 ps-long) or BCF-12 scintillation pulses ignited by a pulsed UV LED (~10 ns-long). Therefore, the statements from [5] on the nature of the pulse shape distortion are not fully applicable for the large-size EQR20 11-6060D-S SiPMs.

Shortening of the rise time could be caused by an abrupt increase in the local electric field strength during the avalanche, and shall thus depend on the pulse charge rather than on the number of microcells fired. Such dependencies measured at various overvoltage follow each other most closely, as shown in Fig. 16 (top-right) – the rise time shortens twofold for signals as large as 2.7 nC with a full spread of ±18% (or 14% s.d.).Definitive explanation of the nature of the pulse shape distortion requires precise info on the internal structure of the NDL SiPMs, which is not public. Although, as shown in Fig.16, NDL EQR20 11-6060D-S SiPMs maintain the ~1:1000 dynamic range unaffected by any kind of distortion (from single photon up to ~100 pC). We assume this range is a subject of optimization of the internal structure of the EQR SiPMs.

It is worth to note that variation of the rise time may not be pronounced in case the input light pulse is longer than the rise time of the undistorted pulse (~3 ns), which is the case for the BCF-12 scintillation fibers projected for the ALICE FoCal-H hadron calorimeter (see below). While this and other key performance parameters of the NDL EQR20 11-6060D-S SiPMs allow for their application as the photon readout option in FoCal-H, elongation of the recovery time as a function of the microcell occupancy poses a significant complication for signal processing and detector calibration.

Apart from a Poisson-based correction of the signal amplitude applicable to any SiPM type, NDL EQR20 11-6060D-S SiPMs require additional correction to reconstruct the linear dependence between the input pulse intensity and output signal amplitude. Moreover, the distortion of the signal shape affects the results of time-over-threshold (TOT) measurement. Extension of the pulse tail not only reduce the TOT dynamic range, but, depending on the threshold used, can create ambiguity in signal charge reconstruction from TOT readings if a coupling capacitance is used. See Fig.18 for the comparison of pulse shapes in log scale measured with the coupling capacitance C2 shorted (top-left) and with C2 = 62 pF (top-right). Waveforms in the bottom plot are also measured with C2 = 62 pF but under the BCF-12 scintillation light, representing the realistic pulses one can see with the NDL EQR20 11-6060D-S SiPM from a scintillator-based hadron calorimeter.

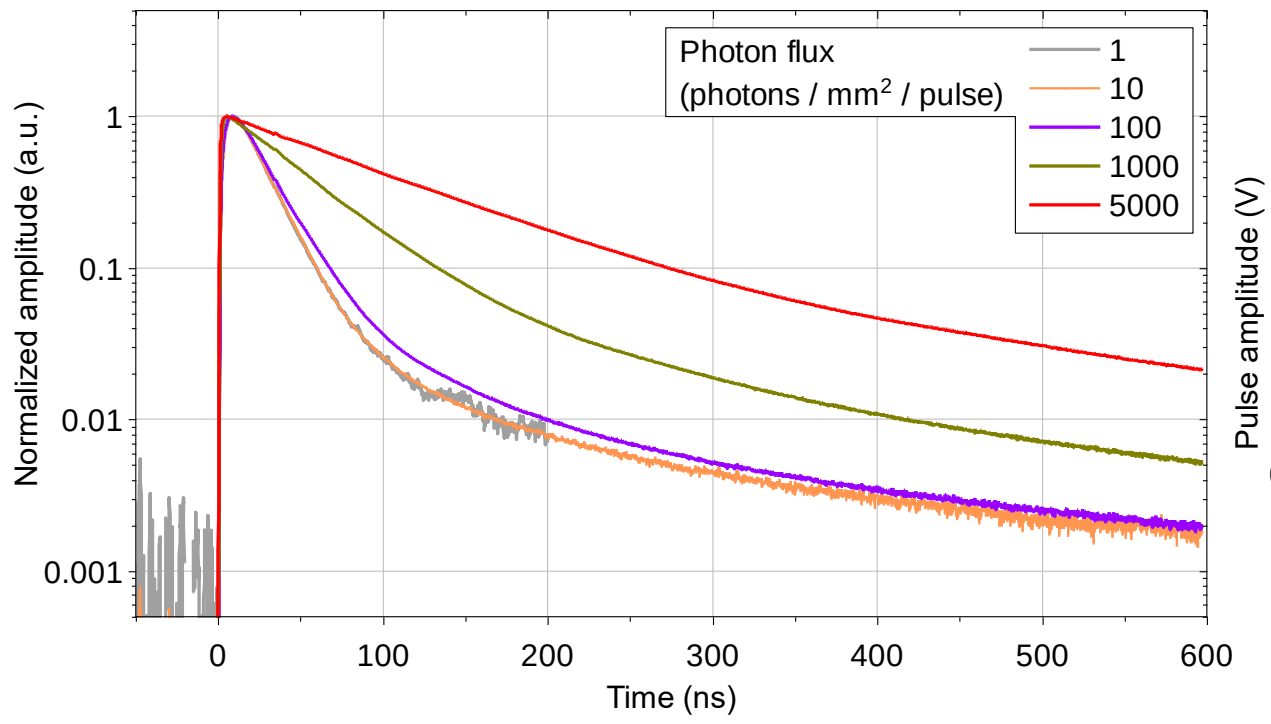

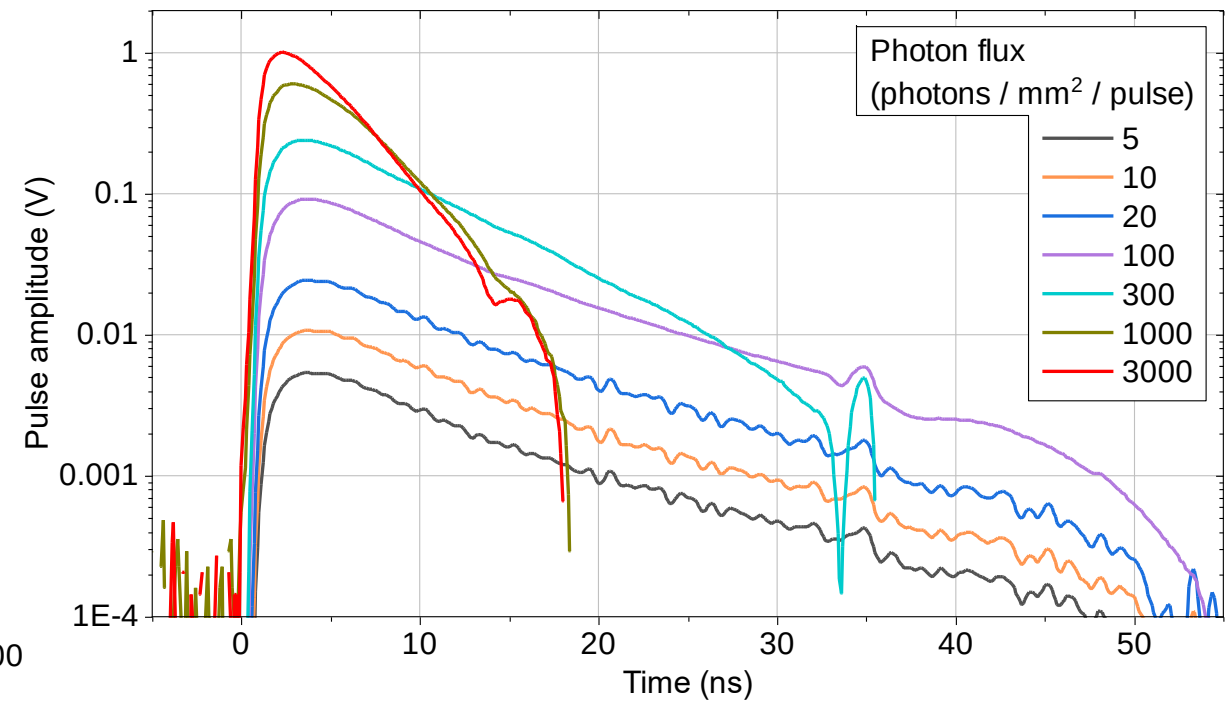

Figure 19. Comparison of the NDL EQR20 11-6060D-S SiPM pulse shapes (log scale) measured under picosecond laser illumination with the coupling capacitance C2 shorted (top-left) and with C2 = 62 pF (top-right). Waveforms in the bottom plot are also measured with C2 = 62 pF, but under the BCF-12 scintillation light (~10 ns-long pulses).

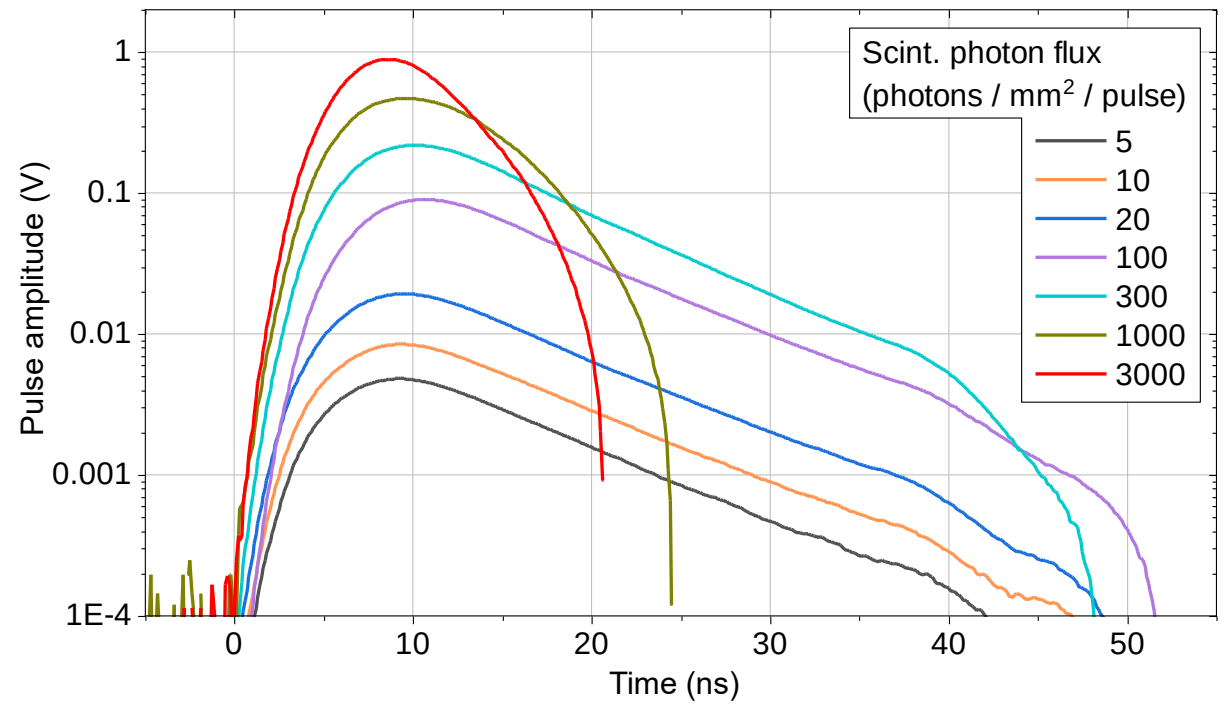

As could be seen from Fig.14 (left) and Fig. 15 (left), rising edge of the NDL pulse varies from ~3 ns down to ~1 ns as the result of the distortion. It therefore can heavily affect the results of time-of-arrival (TOA) measurement for the case of detecting light flashes shorter than that. However, the decay time of most scintillators is larger, including the BCF-12 scintillation fibers used to measure the pulse shapes presented in Fig.18 (bottom). In that case, the NDL pulse shape distortion affects TOT readings only, without introducing a significant TOA error.

The possible advantage of the EQR20 11-6060D-S SiPMs in terms of the effect of radiation damage discussed in Section 1 becomes unjustified at higher dark count rates since many pixels will be "fired" at any given time, therefore increasing the microcell occupancy and widening the pulse shape for both signal and noise pulses.

## 5. Conclusions

NDL EQR20 11-6060D-S SiPMs feature a set of remarkable performance characteristics independently determined by our study. A PDE as high as 50% combined with the large sensitive area (39 mm$^2$), gain of up to $10^6$, and pulses as narrow as 1.5 mV/pC can make them the photosensors of choice in specific applications like, e.g., ring imaging Cherenkov detectors. Mediocre levels of correlated noise and non-ideal stability of the pulse shape at various overvoltages are among the non-critical disadvantages of the EQR technology.

However, the strong pulse shape distortion at higher occupancies will overly complicate the use of the NDL EQR20 11-6060D-S SiPMs in detection of signals of a wide dynamic range that are expected in the ALICE FoCal-H scintillation fiber-based high-energy hadron calorimeter. This effect includes shortening of the rising edge and elongation of the trailing edge. The former effect may be related to an abrupt enhancement of local electric field strength during the avalanche, while the nature of the latter effect is less clear.

Rising edge of signals from the NDL EQR20 11-6060D-S SiPMs is free from distortion up to 100 pC of charge and the falling edge saturates at occupancies above ~2%. The device response is therefore fully linear within a 1:1000 dynamic range, starting from single photon level. Modification of the internal structure of this particular SiPM type can further widen this range, paving the way of the broader use of EQR SiPMs.

**Acknowledgement**

I. Bearden, S. Jia, I. P. Møller and the ALICE FoCal-H project, through which all the studied SiPMs were acquired, gratefully acknowledge support from The Carlsberg Foundation (CF21-0606) and the Danish Council for Independent Research/Natural Sciences.

Sofia University acknowledges support by the European Union - NextGenerationEU through the National Recovery and Resilience Plan of the Republic of Bulgaria, project SUMMIT BG-RRP-2.004-0008-C01. V. Buchakchiev acknowledges support from ESA through contract number 4000142764/23/NL/MH/rp.